\documentstyle[12pt,axodraw]{article}
\newcommand{\NP}{Nucl. Phys. }
\newcommand{\PR}{Phys. Rev. }
\newcommand{\PRL}{Phys. Rev. Lett. }
\newcommand{\PL}{Phys. Lett. }

\begin{document}

\pagenumbering{arabic}

\begin{flushright}
TUIMP-TH-99/107

AS-ITP-99-13
\end{flushright}

\begin{center}
{\Large\sf Enhanced contribution to quark and neutron electric dipole
moments with small mixing of right-handed currents and CKM CP violation}
\\[10pt]
\vspace{1.0 cm}

{Yi Liao}
\vspace{1.5ex}

{\small Department of Modern Applied Physics, Tsinghua University,
Beijing 100084, P.R.China\\}
\vspace{3.0ex}
{Xiaoyuan Li}
\vspace{1.5ex}

{\small Institute of Theoretical Physics, The Chinese Academy of Sciences,
Beijing 100080, P.R.China\\}
\vspace{3.0ex}
{\bf Abstract}
\end{center}

We study the light quark and the neutron electric dipole moments (EDMs)
under the assumptions that the CP source is still in the usual CKM matrix
and that there is a small mixing of right-handed charged currents in the
quark sector. We find that the EDMs arise already at two loop order that
are much larger than the standard model (SM) result even for a small
mixing.

\begin{flushleft}
\end{flushleft}

\begin{flushleft}
\end{flushleft}

\newpage

One of the outstanding questions that remain to be answered in nowadays
particle physics is the origin of CP violation$\cite{review}$. Even
after thirty five years since its discovery our knowledge is still
mainly limited to the neutral kaon system. However, to achive a
consistent picture of CP violation, it is extremely desirable to
observe its effects in other systems. In relativistic quantum field
theories, a CP violating interaction can induce P and T violating
electric dipole moments (EDMs) for elementary particles. So, one
promising possibility beyond the kaon system is provided by the
observation of the neutron and the electron EDMs. The current
experimental bounds on them are respectively,
$|d(e)|< 4.3\times 10^{-27}{\rm e~cm}$$\cite{electron}$,
$|d(n)|< 1.1\times 10^{-25}{\rm e~cm}$$\cite{neutron}$.
Recently, the bound on $d(n)$ has been pushed to
$6.3\times 10^{-26}{\rm e~cm}$$\cite{recent}$ and significant
improvement is expected to be available in the near future
$\cite{future}$.

Due to its fundamental importance and experimental progress there has
been continuing theoretic interest in searching for mechanisms to
induce EDMs. In the standard model (SM) of electroweak interactions,
the only CP violating parameter is the CP phase in the CKM matrix of
the quark left-handed charged currents$\cite{km}$. There is no
contribution to quark ( and thus neutron ) EDMs at one loop order
because the relevant
amplitude is proportional to the moduli of the CKM matrix. Naively,
the first contribution should come from the two loop amplitude
that is rich enough in the flavour structure. However, as first pointed
out by Shabalin$\cite{shab}$ and confirmed afterwards by others
$\cite{donoghue}$-$\cite{ll99}$, the two loop contribution
to quark EDMs actually vanishes strictly. The same null result was also
witnessed in the $W^{\pm}$ EDM$\cite{kp91}\cite{booth}$. This is
surprising in the sense that there is no symmetry which would demand
the vanishing of EDMs at the lowest non-trivial order. Actually, when
QCD is turned on, quark EDMs already arise at three loop order
$\cite{shab80}$-$\cite{ck97}$. The numerical result is indeed too small
to be observable in the near future. It seems therefore that the CKM
mechanism of CP violation would be irrelevant to experimental search
for EDMs although there has been no clear reason why this must be so.
Beyond the SM, many new possibilities are open to CP violation.
Generally the EDMs are induced at one loop order that usually turn out
to be too large. As a result, some degree of fine-tuning is necessary
to suppress these contributions. For instance, in the minimal
supersymmetric standard model, one has to appeal to assumptions about
the size of CP phases$\cite{size}$ and the spectrum$\cite{mass}$, or
the cancellation among different contributions$\cite{cancel}$.
In this regard, EDMs should most naturally arise at two loop order.

In this note, we shall stick to the more conservative possibility that
CP violation is still restricted to be in the CKM matrix as in the SM
but a small mixing of right-handed charged currents is allowed in the
quark sector. This consideration is partly motivated by the recent
result on $\epsilon^{\prime}/\epsilon$$\cite{kTeV}$ which indicates
that we should surely take the CKM mechanism seriously. We would like
to investigate how far we can go with only CKM CP violation. We observe
that with mixed left- and right-handed charged currents the chirality
flip required by the EDM operator is not necessarily to be implemented
by a small external quark mass but could be done by an internal quark
mass. It is then possible to obtain terms that are not suppressed by
external quark masses and thus gain an enhancement in EDMs as compared
to the SM case. Theoretically, the model considered here may arise,
for example, as a special case of left-right symmetric models
$\cite{lrmodel}$ in which the only CP source is assumed to be in the
usual CKM matrix. These models have received heated attention
$\cite{babu}$ due to recent experimental evidences for neutrino
oscillations$\cite{fukuda}$ and a possible anomaly in the
$Z \rightarrow b \bar{b}$ decay$\cite{marciano}$ .

We shall consider the light quark EDMs from which the neutron EDM is
constructed by using the $SU(6)$ relation,
\begin{equation}
\displaystyle d(n)=\frac{4}{3}d(d)-\frac{1}{3}d(u).
\end{equation}
Possible enhancements from long distance physics will not be considered
here$\cite{distance}$. The effective Lagrangian for the EDM interaction
is defined as
\begin{equation}
\displaystyle {\cal L}_{\rm eff}=
-\frac{i}{2}d~\bar{\psi}\gamma_5\sigma_{\mu\nu}\psi F^{\mu\nu},
\end{equation}
where $F^{\mu\nu}$ is the electromagnetic tensor and $d$ is the EDM of
the fermion $\psi$. As mentioned previously, the CP violating
interaction is assumed to be ,
\begin{equation}
\displaystyle {\cal L}_W=
\frac{g}{\sqrt{2}}\sum_{\alpha,i}V_{\alpha i}\bar{u}_{\alpha}\gamma^{\mu}
(P_L c_{\theta}+P_R s_{\theta})d_i W^+_{\mu}+{\rm h.c.}.
\end{equation}
Here
$\displaystyle P_L=\frac{1}{2}(1-\gamma_5),~P_R=\frac{1}{2}(1+\gamma_5)$
and the Greek and Latin letters denote respectively the up- and down-type
quarks. $V_{\alpha i}$ is the entry $(\alpha,i)$ of the CKM matrix.
$c_{\theta}=\cos\theta,~s_{\theta}=\sin\theta$ where $\theta$ is the
mixing angle of the left-right gauge bosons $W^{\pm}_{L,R}$, and
$W^{\pm}$ is to be identified as the usual charged weak boson. The
mixing angle is constrained by low energy processes to be
$\displaystyle |\theta|<10^{-2}\sim 10^{-3}$$\cite{pdg}$.
We stress again that we have assumed that the only CP source is the
single phase in the usual CKM matrix. If this is not the case, EDMs
will generally be induced at one loop order due to the mismatch between
two CKM matrices in the left- and right-handed currents and due to the
new phase in the $W^{\pm}_{L,R}$ mixing$\cite{lrone}$$\cite{lrmore}$.
We have also safely neglected contributions from extra heavier charged
gauge bosons. Since we shall work in renormalizable gauges, we need to
specify the interaction Lagrangian for the would-be Goldstone bosons
$G^{\pm}$. The easiest way to do so is by examining the cancellation
of the gauge parameter dependence between $W^{\pm}$ and $G^{\pm}$
contributions to a physical process, e.g.,
$\bar{u}_1d_1\to\bar{u}_2d_2$. We find,
\begin{equation}
\displaystyle
{\cal L}_G=\frac{g}{\sqrt{2}}\frac{1}{m_W}\sum_{\alpha,i}
V_{\alpha i}\bar{u}_{\alpha}\left[c_{\theta}(m_{\alpha}P_L-m_i P_R)+
s_{\theta}(m_{\alpha}P_R-m_iP_L)\right]d_iG^+ +{\rm h.c.}.
\end{equation}
To simplify the computation of diagrams involving $W^{\pm}$ exchange,
we shall use the background field$\cite{abbott}$-$\cite{ll95}$
( or the nonlinear$\cite{nonlinear}$ ) $R_{\xi}$ gauge with $\xi=1$.
The relevant $W^+W^-A$ vertex ( $A$ is the background electromagnetic
field ) is given in Ref.$\cite{ll99}$

The Feynman diagram that contributes to the quark EDMs is shown in
Fig.1. There are four groups of contributions, namely $WG,~GW,~WW$ and
$GG$, where the first and second letters refer to the bosons exchanged
in the outer and inner loops respectively. The final $\gamma_5$ in the
effective EDM operator can only come from vertices since there would be
no P violation if no $\gamma_5$ were involved in these vertices.
Details concerning renormalization and calculational techniques will
not be presented here. We refer the interested reader to
Ref.$\cite{ll99}$ although completely new terms will appear in the
current case as a consequence of the left-right mixing. In that work,
we studied possible contributions to the quark EDMs arising from charged
Higgs bosons in the two Higgs doublet model with only CKM CP violation,
\begin{equation}
\displaystyle {\cal L}_H=\frac{g}{\sqrt{2}}\frac{1}{m_W}\sum_{\alpha,i}
V_{\alpha i}\bar{u}_{\alpha}(C_{\alpha i}+C^{\prime}_{\alpha i}\gamma_5)
d_iH^+ +{\rm h.c.},
\end{equation}
where $C_{\alpha i}$ and $C^{\prime}_{\alpha i}$ are real coulpings
depending on the masses of $u_{\alpha}$ and $d_i$. We found that the
contribution to the EDM of the quark $u_e$ has the separate form,
\begin{equation}
\displaystyle d(u_e)={\rm Im}(V_{ek}V^*_{\alpha k}V_{\alpha j}V^*_{ej})
[H(m_k)-H(m_j)],
\end{equation}
if $C_{\alpha i}$ and $C^{\prime}_{\alpha i}$ are related by the
following relations,
\begin{equation}
\displaystyle
C_{\alpha i}=xm_i+ym_{\alpha},~C^{\prime}_{\alpha i}=xm_i-ym_{\alpha}.
\end{equation}
Here $H$ is also a function of masses of bosons, $u_{\alpha}$ and $u_e$,
but the crucial point is that it depends exclusively on $m_k$ or $m_j$.
$x$ and $y$ are mass-independent constants that depend on the detail
of the model. It is this separate structure that leads to the complete
cancellation when summation over internal flavours is taken and when
the unitarity of CKM matrix is assumed. We can see from Eqn.$(4)$ that
the above relations are not respected any longer when the left-right
mixing is introduced. Therefore, we shall probably obtain a non-vanishing
result in the current case.

Our result should depend on $c_{\theta}$ and $s_{\theta}$ to the fourth
power. If the charged current is purely left-handed ($s_{\theta}=0$),
we return to the vanishing result in the SM. The purely right-handed
case ($c_{\theta}=0$) is related by a parity reflection, i.e., by
reversing the sign of EDM. If the current is purely vectorial
($c_{\theta}=s_{\theta}$) or purely axial-vectorial
($c_{\theta}=-s_{\theta}$), we still arrive at a vanishing
result. Therefore, the final, possibly non-vanishing result should be
proportional to $(c_{\theta}^2-s_{\theta}^2)c_{\theta}s_{\theta}$. Let
us now examine qualitatively what quark mass dependence should be
expected. The CP source is in the CKM matrix which arises from
diagonalization of quark mass matrices. If any two up-type or down-type
quarks are degenerate, there will be no CP violation and thus no EDMs.
This implies, for example, that the contribution to $d(u_e)$ from Fig.1
must be antisymmetric in $m_j$ and $m_k$. Hence the pair of diagrams
related by mirror reflection $j\leftrightarrow k$ contribute the same
to $d(u_e)$ which is proportional to
${\rm Im}(V_{ek}V^*_{\alpha k}V_{\alpha j}V^*_{ej})$ while the
${\rm Re}(V_{ek}V^*_{\alpha k}V_{\alpha j}V^*_{ej})$ part is cancelled
as required by Hermiticity$\cite{ck}$. To proceed further, we make use
of the hierarchical structure in quark masses,
$m_t\gg m_W\gg m_b\gg m_c\gg m_s\gg m_{u,d}$.
At the moment, all quarks except the top are treated as light on the
same footing as compared to the reference scale $m_W$. We shall be
concerned with the first non-trivial terms which are of lowest order
in light quark masses. Higher order terms are safely ignored even if
they are possibly enhanced by factors of $m_t^2$ since
$\displaystyle\frac{m_q^2}{m_W^2}\frac{m_t^2}{m_W^2}\ll 1$,
where $q$ can be any quark except the top. The external quark mass
$m_u$ or $m_d$ is also ignored from the beginning. Then, the chirality
flip is to be made by internal quark masses so that the result depends
on them to some odd power. We first consider the contribution to the EDM
of the $u$ quark, $d(u)$. In this case, $u_{\alpha}$ can be heavy ($t$)
or light ($c$) but $d_j$ and $d_k$ are always light. Due to the
unitarity of the CKM matrix, the desired terms must involve $d_j$ and
$d_k$ masses simultaneously in order to survive the summation over
the flavour pair $(jk)$. Therefore, we expect
$d(u)\propto m_jm_k(m_j-m_k)$ up to logarithms associated with each
term. For the EDM of the $d$ quark, we should first discriminate two
cases; namely, the quarks $u_{\alpha}$ and $u_{\beta}$ are both
light ($u,~c$) or one of them is heavy ($t$). Since $d_i$ is always
light, the desired terms must involve its mass; otherwise they will
be cancelled upon summing over the flavour $i$. Therefore, if
$u_{\alpha}$ and $u_{\beta}$ are both light as well, we expect
this part of contribution to be proportional to
$m_{\alpha}m_{\beta}(m_{\alpha}-m_{\beta})m_i^2$
so that it can be ignored. Now suppose $u_{\alpha}$ is the heavy top.
Although the amplitude is originally antisymmetric with respect to
$m_{\alpha}$ and $m_{\beta}$, this antisymmetry is not preserved by
the expansion according to the hierarchy $m_t\gg m_W\gg m_q$.
Similar arguments then indicate that the surviving result must involve
$m_{\beta}$ so that we expect $d(d)\propto m_{\beta}m_i^2$.

The above analysis gives no information about possible enhancements by
factors of $m_t^2$ or logarithms. An explicit calculation is therefore
necessary. The relevant amplitude consists of two sets of terms, one
proportional to
$(c_{\theta}^2-s_{\theta}^2)(c_{\theta}^2+s_{\theta}^2)$, the other
proportional to
$(c_{\theta}^2-s_{\theta}^2)c_{\theta}s_{\theta}$. The first set is of
the separate form as shown in Eqn.$(6)$ and cancelled by the unitarity
of the CKM matrix upon summing over the flavour pair $(jk)$. One or
two terms in the second set are accidentally of the separate form and
thus also cancelled, but more others are not. These latter terms
produce the following leading contributions which are least suppressed
by light quark masses,
\begin{equation}
\begin{array}{rcl}
\displaystyle
d(u)&=&+\frac{eG_F^2}{32\pi^4}\tilde{\delta}~m_b^2m_s
(c_{\theta}^2-s_{\theta}^2)c_{\theta}s_{\theta}\\
\\
\displaystyle
&&\left\{\left[\left(\frac{43}{18}Q_d\right)+0+0+
\left(Q_u\left(\frac{1}{24}+\frac{1}{4}\ln\frac{m_t^2}{m_W^2}\right)
+Q_d\left(-\frac{5}{72}-\frac{1}{6}\ln\frac{m_t^2}{m_W^2}\right)\right)
\right]\right.\\
\\
\displaystyle
&&\left.-\left[\left(Q_u\left(-\frac{5}{2}+\frac{\pi^2}{3}\right)-Q_d\right)+
\left(-\frac{5}{4}Q_u+\frac{3}{2}Q_d\right)+
\left(Q_u\left(-\frac{19}{3}+\frac{2\pi^2}{3}\right)-\frac{1}{3}Q_d\right)+0
\right]\right\}\\
\\
\displaystyle
&=&+\frac{eG_F^2}{32\pi^4}\tilde{\delta}~m_b^2m_s
(c_{\theta}^2-s_{\theta}^2)c_{\theta}s_{\theta}
\left[Q_u\left(\frac{81}{8}-\pi^2+\frac{1}{4}\ln\frac{m_t^2}{m_W^2}\right)+
Q_d\left(\frac{155}{72}-\frac{1}{6}\ln\frac{m_t^2}{m_W^2}\right)\right],\\
\\
\displaystyle
d(d)&=&-\frac{eG_F^2}{32\pi^4}\tilde{\delta}~m_b^2m_c
(c_{\theta}^2-s_{\theta}^2)c_{\theta}s_{\theta}\\
\\
\displaystyle
&&\left[\left(-4Q_u+3Q_d\right)+
\left(Q_u\left(5-\frac{2\pi^2}{3}\right)+
Q_d\left(-\frac{1}{2}+\frac{\pi^2}{6}+\frac{1}{2}\ln\frac{m_b^2}{m_W^2}
\right)\right)\right.\\
\\
\displaystyle
&&~\left.+0+\left(Q_u\left(-\frac{1}{2}+\frac{\pi^2}{6}-\frac{3}{4}
\ln\frac{m_t^2}{m_W^2}\right)+Q_d\left(-\frac{5}{4}+\frac{3}{4}
\ln\frac{m_t^2}{m_W^2}\right)\right)\right]\\
\\
\displaystyle
&=&-\frac{eG_F^2}{32\pi^4}\tilde{\delta}~m_b^2m_c
(c_{\theta}^2-s_{\theta}^2)c_{\theta}s_{\theta}\\
\\
\displaystyle
&&
\left[Q_u\left(\frac{1}{2}-\frac{\pi^2}{2}-\frac{3}{4}
\ln\frac{m_t^2}{m_W^2}\right)+Q_d\left(\frac{5}{4}+\frac{\pi^2}{6}+
\frac{3}{4}\ln\frac{m_t^2}{m_W^2}+\frac{1}{2}\ln\frac{m_b^2}{m_W^2}\right)
\right].
\end{array}
\end{equation}
Some explanations are in order. $\tilde{\delta}$ is the usual rephasing
invariant of CP violation$\cite{jarlskog}$,
which is $c_1c_2c_3s_1^2s_2s_3\sin\delta$ in
the original parameterization of the CKM matrix$\cite{km}$.
In the above formulas, only the largest terms are kept. The two terms
in the braces for $d(u)$ originate from the top and the charm quarks
respectively; $d(d)$ is contributed totally by the top quark. The four
terms in each square brackets of the first equalities for $d(u)$ and
$d(d)$ arise from $WG,~GW,~WW$ and $GG$ exchanges in loops. We found
no leading terms which are enhanced by $m_t^2$ or di-logarithms. The
absence of $m_t^2$ enhancement is consistent with general arguments
based on gauge invariance and naive dimensional analysis $\cite{lkl}$.
For numerical analysis, we take the following input parameters:
$G_F\sim 1.2\times 10^{-5}{\rm ~GeV}^{-2},
~\tilde{\delta}\sim 5\times 10^{-5},
~m_t\sim 175{\rm ~GeV},~m_b\sim 4.5{\rm ~GeV},~m_c\sim 1.5{\rm ~GeV},
~m_s\sim 200{\rm ~MeV},~m_W\sim 80{\rm ~GeV}$.
We obtain,
\begin{equation}
\begin{array}{l}
\displaystyle
d(u)\sim-4\times 10^{-34}\frac{s_{\theta}}{10^{-2}}{\rm ~e~cm},~~
d(d)\sim+6\times 10^{-32}\frac{s_{\theta}}{10^{-2}}{\rm ~e~cm},\\
\displaystyle
d(n)\sim+8\times 10^{-32}\frac{s_{\theta}}{10^{-2}}{\rm ~e~cm}.
\end{array}
\end{equation}
These numbers are generally too small to be observable in the near
future if there is no significant enhancement from long distance
physics, but they are still much larger than the values in the SM
$\cite{ck97}$,
\begin{equation}
\displaystyle
d(u)\sim-2\times 10^{-35}\frac{m_u}{5{\rm ~MeV}}{\rm ~e~cm},~~
d(d)\sim-8\times 10^{-35}\frac{m_d}{10{\rm ~MeV}}{\rm ~e~cm}.
\end{equation}

To summarize, we have considered the light quark and the neutron EDMs
under the assumptions that the CP source is still in the usual CKM
matrix and that there is a small mixing of right-handed charged
currents in the quark sector. We found that the EDMs arise already at
two loop order. The effect is of normal size in the sense that compared
to new physics effects studied in the literature, it is small only
because the CP parameter $\tilde{\delta}$ and the mixing angle $\theta$
are most stringently constrained in the current case, which bring
down EDMs by a factor of $10^7$. The enhancement over the SM is also
significant. This phenomenon caused by the left-right mixing was also
observed in the decay $b\to s\gamma$$\cite{bsgamma}$.
Finally our study provides a good explanation of the result
in the SM. There are two points which are responsible for the vanishing
result at two loop order. One is the unitarity of the CKM matrix, the
other is the pure chirality structure of charged currents. This pure
chirality strucure diminishes further the non-vanishing result at three
loop order by introducing a small $u$ or $d$ quark mass to flip the
chirality of the EDM operator.

\vspace{0.5cm}

Y.L. would like to thank the High Energy Physics Group of Michigan
State University for its hospitality where part of the work was
done during a visit. He is grateful to C.-P.Yuan for his interest
and helpful discussions.

\vspace{0.5cm}

\begin{flushleft}
{\large Figure Captions }
\end{flushleft}

\noindent
Fig. 1  The Feynman diagram that contributes to the EDM of the up-type
quark $u_e$. A background electromagnetic field is understood to be
attached to internal lines in all possible ways. The dashed lines
represent $W^{\pm}$ and $G^{\pm}$ bosons. The diagram for the down-type
quark $d_e$ is obtained by substitutions:
$\alpha\to i$ and $j,~k\to\alpha,~\beta$.

\newpage
\newpage
\begin{center}
\begin{picture}(200,600)(0,0)
\SetOffset(10,80)\SetWidth{2.}

\SetOffset(0,300)
\ArrowLine(20,100)(45,100)
\ArrowLine(45,100)(70,100)
\ArrowLine(70,100)(130,100)
\ArrowLine(130,100)(155,100)
\ArrowLine(155,100)(180,100)
\DashCArc(100,100)(55,0,180){4.}
\DashCArc(100,100)(30,0,180){4.}
\Text(33,111)[]{$e$}
\Text(58,111)[]{$j$}
\Text(100,111)[]{$\alpha$}
\Text(143,111)[]{$k$}
\Text(168,111)[]{$e$}
\Text(58,89)[]{$(\alpha)$}
\Text(100,89)[]{$(i)$}
\Text(143,89)[]{$(\beta)$}
\SetOffset(0,-400)

\SetOffset(100,200)
\Text(5,5)[]{\large Figure $1$}

\end{picture}\\
\end{center}

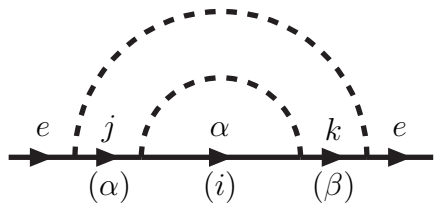
\end{document}